\documentclass[manuscript,screen,nonacm]{acmart}

\setcopyright{none}
\renewcommand\footnotetextcopyrightpermission[1]{}

\usepackage{longtable}
\usepackage{float}
\begin{document}

\title{Beyond accuracy: what AI assurance needs to establish in public services}
\author{Louise Yorke}
\affiliation{%
  \institution{University College London}
  \city{London}
  \country{United Kingdom}}
\begin{abstract}
Public bodies increasingly rely on AI-produced evidence, including classifications, summaries and assessments. These outputs can inform policy decisions and service delivery, making their reliability a matter of public accountability. Establishing whether this evidence can be relied upon requires testing its outputs and examining how organisations decide whether those tests are sufficient. We combine technical evaluation with analysis of AI governance requirements, informed by measurement theory and management-based regulation.

We compare eight GPT-4o configurations classifying 1,200 UK Prime Minister's Questions into 21 policy topics, against human-coded labels and across unchanged repeats. Accuracy ranges from 68.2\% to 70.3\%, yet 24.3\% of questions receive different labels, with changes in topic shares and rankings. Configurations agree more with one another than with human coding: agreement does not guarantee correctness. Under the implemented score-extraction procedure, the proportion flagged for human review ranges from 0.9\% to 41.0\%.

We analyse 37 UK and EU policy documents. Both sets expect testing and oversight, but reporting expectations do not consistently cover changing answers, shared errors, selective evaluation, substantive conclusions or review routing. Some UK guidance explicitly addresses stability. The UK sample emphasises public bodies checking their own use; the EU places substantial testing duties on providers while leaving them important choices about measures and acceptance thresholds.

Without appropriate checks, evidence could misrepresent policy issues, redirect attention or resources, and leave errors undetected. Public policy should require organisations to justify reliance through checks beyond aggregate accuracy, clear responsibility, access to expertise and scrutiny of results.
\end{abstract}



\maketitle

\section{Introduction}

Public bodies increasingly use AI to produce the classifications, summaries, assessments and advice through which they understand situations and conduct their work. The European Commission's AI@EC communication reports using AI to locate scientific evidence for policymaking and identifies further opportunities for analytical and drafting support~\cite{ec2024aiatec}. The UK's Consult transparency record describes a tool under development that extracts themes from consultation responses and presents them through dashboards for policymakers~\cite{dsit2025consult}.

AI can influence public decisions through the evidence it produces, even when it does not determine an individual's entitlement or make the final decision. What a system includes, how it groups material and which patterns it makes prominent can shape the information available to decision-makers. Variation may be acceptable in exploratory work while requiring closer examination when outputs support service assessments, resource allocation, policy priorities or public scrutiny. In these settings, differences in how evidence is classified or summarised could direct attention towards some concerns at the expense of others. Keeping a human responsible for the final decision does not resolve an earlier question: what justifies relying on the account the system produces?

Answering this question requires both technical testing and examination of how organisations justify their reliance on the results. Tests can show how outputs behave, but they cannot alone determine what variation is acceptable for a particular use, who should take responsibility for that judgement or how it can be challenged. Governance requirements can specify these responsibilities, but their existence does not establish that outputs are dependable or that checks work in practice.

Assurance addresses many properties, including safety, fairness, security and performance~\cite{UK206}.
It includes technical testing alongside documentation, organisational procedures, review and audit. This paper examines three properties relevant to reliance on AI-produced information: stability across repeated runs and reasonable configurations, correctness against a reference, and whether the conclusions drawn from the outputs remain robust. These properties are connected but distinct. Similar aggregate performance can accompany conflicting classifications of individual cases, while agreement between configurations need not establish correctness. Conversely, differences between individual classifications may leave a broad conclusion intact~\cite{jacobs2021measurement,marx2020predictive}.

We ask: \emph{How do UK and EU assurance requirements provide for assessing and justifying the dependability of AI-produced information in public services?} First, using GPT-4o to classify 1,200 UK Prime Minister's Questions into 21 policy topics, we compare outputs across eight configurations, against human-coded labels and across five unchanged repeats. The experimental findings then informed what we looked for in 37 UK and EU documents specifying assurance requirements: how outputs should be checked, who should carry out those checks, who sets the standards, what support is available, who can scrutinise the assessment and what should happen when problems are found.

\begin{description}
  \item[RQ1] How do repeated runs and reasonable configuration choices affect the stability and correctness of individual classifications?
  \item[RQ2] What do these differences change about the substantive account, reported performance and implied human-review requirements?
  \item[RQ3] What do UK and EU requirements ask organisations to establish about AI outputs, and how are the expected checks and standards specified?
  \item[RQ4] How do these requirements allocate standard-setting, checking, support, scrutiny and action on results?
\end{description}

The paper makes three contributions. First, it shows that configurations of the same model with similar accuracy assign different labels to about a quarter of questions. It also shows how excluding disputed questions raises reported accuracy and how the implemented score-extraction procedure changes the proportion flagged for human review. Second, it shows that UK and EU requirements expect testing and oversight while leaving important choices about measures, thresholds and implementation to providers and public bodies. Third, it demonstrates why discretion over evaluation requires justification: choices about configuration, filtering and review procedures can change what an assessment appears to establish.

The next section reviews the related research and develops the theoretical basis for the analysis, drawing on measurement theory and management-based regulation.

\section{Related work and theoretical framework}

Research on LLM evaluation and human oversight shows that outputs can vary across prompts, runs and configurations even when aggregate performance is similar, and that confidence scores and human review do not by themselves establish that outputs can be relied on. Our paper extends this work by examining how UK and EU requirements address the assessment of AI-produced evidence and allocate responsibility for justifying reliance on it. We review this research and then set out the theoretical framework, drawing on measurement and management-based regulation.

\subsection{Stability, correctness and multiplicity in LLM classification}

\citet{mizrahi2024state} demonstrate substantial variation in performance across instruction templates and recommend evaluation across multiple prompts. \citet{barrie2026prompt} develop a framework for assessing prompt stability through agreement measures adapted from annotation research. \citet{yorke2026behind} also examines variation across repeated runs in LLM-as-judge evaluation.

Predictive multiplicity provides a related reason to look beyond aggregate performance. \citet{marx2020predictive} show that competing classifiers with similar performance can assign conflicting predictions to individual cases. \citet{damour2022underspecification} show that predictors with similarly strong test performance can behave differently in deployment settings. This work motivates examining whether similar performance conceals differences in the evidence produced. We therefore examine how variation affects individual classifications, shared errors, the substantive account constructed from labels and an implemented procedure for routing outputs to review. Our experiment examines operational configurations of an LLM classification pipeline rather than the model-selection problem investigated in the multiplicity studies.

\subsection{Confidence scores and human review}

Public-sector research also questions whether formal oversight arrangements establish that outputs can be used appropriately. \citet{green2022flaws} argues that policies requiring human oversight frequently rely on assumptions about reviewers' capabilities that need empirical support. Research on LLM confidence likewise distinguishes the production of a confidence estimate from its usefulness for calibration or failure prediction~\cite{xiong2024uncertainty}. Confidence comparisons can change with the measurement protocol even when answers and correctness labels are held fixed~\cite{kim2026confidence}, and a score's ranking ability differs from the range of thresholds it makes available for selective prediction~\cite{sun2026granularity}. These concerns motivate scrutiny of how outputs reach reviewers, what information accompanies them and what reviewers are expected to establish.

\subsection{Connecting evaluation and governance}

\citet{frohnapfel2026multiplicity} provide the closest connection between predictive multiplicity and AI assurance requirements. They relate conflicting individual predictions to the AI Act's accuracy provisions and propose measures and information that providers could supply to deployers. Our contribution builds on that connection through a classification experiment and a comparative analysis of UK/EU assurance documents. We examine how output differences affect the evidence produced, alongside how formal requirements organise its assessment and justification. \citet{tao2026checkers} examine AI Act compliance-support tools and find variation in their coverage and the guidance they provide; their study concerns tools that interpret obligations, whereas ours examines formal requirements. The wider concern is consequential reliance on AI-produced information in public services, including uses that shape understanding without determining a final decision.

\subsection{Theoretical framework}

The analysis draws on two frameworks. We treat AI-assisted topic classification as a measurement of policy content. \citet{jacobs2021measurement} argue that measurement depends on assumptions that need to be explicit and tested. They distinguish reliability, concerning consistency under similar inputs and conditions, from validity, concerning whether a measure captures what it is intended to capture. In our experiment, stability concerns agreement across runs and configurations, while correctness means agreement with the human-coded CAP reference; neither alone establishes validity for the intended use. Measurement motivates RQ1's distinction between stability and correctness and RQ2's examination of the account constructed from classifications. It also informs RQ3's questions about what checks are expected to establish.

Management-based regulation provides the institutional framework~\cite{coglianese2003management,coglianese2010meta}. It addresses arrangements that require organisations to develop controls while leaving important operational details to them. Task-specific knowledge can justify delegation; discretion is not itself evidence of failure. Applying this perspective to multifunctional AI, \citet{coglianese2025multifunctionality} argue that heterogeneous applications and risks favour proactive management by developers and users, accompanied by regulatory capacity and continuing scrutiny. This perspective directs attention towards how choices are justified, supported and examined. It informs RQ4's analysis of who chooses standards, performs checking, supplies capability, scrutinises the justification and acts on results.

Together, these perspectives connect what technical tests establish with how organisations justify relying on their results. The documentary coding framework organises the evidence by decision-making, delivery, coordination and accountability, each divided into design and resources; the theoretical perspectives guide its interpretation.
\section{Methods}

The study uses a sequential design combining a classification experiment with documentary analysis. The experiment addresses RQ1 and RQ2, examining variation in AI classifications and its implications for reported performance, the account of political attention and hypothetical human review. These findings informed what we looked for in the policy documents: whether requirements address stability, correctness, changes in substantive conclusions and the operation of review procedures. The documentary analysis addresses RQ3 and RQ4, using the governance coding framework to examine these expectations and how responsibility for checking, support, scrutiny and action is allocated. The integrated analysis brings the findings together to assess how the sampled requirements address the issues revealed by the experiment.

\subsection{Classification experiment}

\subsubsection{Corpus and sample}

The UK Comparative Agendas Project (CAP) dataset contains 9,061 unique Prime Minister's Questions (PMQs) after deduplication by date, question text and major topic~\cite{cap_ukpmq}. It covers 21 policy topics between 1997 and 2008. CAP codes the MP's question, not the Prime Minister's response.

We chose CAP because it provides topic labels assigned by human coders, which give a reference against which model labels can be compared. Accuracy is the share of questions for which the model assigns the same topic as CAP.

The topic classes are imbalanced: Government Operations represents 17.4\% of questions, while Culture, Foreign Trade and Technology each represent less than 1\%. We therefore drew a fixed-seed stratified sample of 1,200 questions: 60 from each of 19 sufficiently large topics, 45 from Foreign Trade and all 15 Culture questions. Each question is weighted by its CAP topic's share of the 9,061-question corpus divided by that topic's share of the sample. This adjusts topic-share estimates for the stratified sampling design.

\subsubsection{Configurations and execution}

We test whether plausible configurations produce the same classifications. Different organisations can reasonably choose different pipeline setups because they balance cost, speed, performance, model access and availability, organisational approval and interpretability. The experiment does not identify an optimal setup. The anchor uses GPT-4o, full category descriptions, an explicit coding rule, six few-shot examples, requested JSON output and temperature-zero decoding. It provides no information about the historical period or the asking MP's party. It was selected as a defensible default rather than the highest-scoring pilot configuration.

The analysis uses the eight configurations for which labels and extracted scores were retained: the anchor, descriptions only, no few-shot examples, reasoning and label, chain-of-thought, time scope, speaker scope and free-text output. These configurations come from a 17-configuration experiment run in April 2026. Raw per-call outputs from that experiment were lost when the cloud environment storing them was deleted; parsed labels and extracted scores survive for the eight configurations analysed here. One free-text output is missing, giving 1,199 complete cases for across-configuration grouping and agreement filtering. The appendix summarises the configuration dimensions (Table~\ref{tab:configuration-dimensions}); full specifications and rendered prompts are available in the repository.

\subsubsection{Repeated runs and comparison measures}

The anchor was repeated five times over all 1,200 questions on 1 September 2026, with the prompt, model identifier, question order and decoding settings unchanged. All 6,000 calls parsed successfully. Provider aliases were used rather than dated model snapshots. Because these repeats were conducted about five months after the original experiment, their comparison with the April results cannot separate configuration effects from changes over time.

We compare each configuration with CAP's human-coded labels to assess correctness against that reference, and compare configurations with one another to assess stability. Mean disagreement between pairs of the five unchanged repeats gives a baseline for run-to-run variation. Pairwise AI-AI and AI-CAP agreement use the same 1,199 complete cases.

We report accuracy and error destinations. We compare paired accuracy using exact McNemar tests, with Holm and Benjamini-Hochberg corrections for 28 comparisons. Weighted full-period topic shares allow comparison of topic magnitudes and rankings.

\subsubsection{Agreement filtering and hypothetical review routing}

We group questions by whether all, some or none of the configurations agree with CAP, and split those that never agree by whether their labels match each other. Agreement filtering keeps only questions given the same label by all eight configurations. We compare accuracy before and after filtering, starting from the same 1,199 complete cases.

The extracted score is the highest probability among valid CAP codes at a selected token. It is not necessarily the probability of the final label, and it is not a calibrated probability of correctness. Where reasoning or free text precedes the answer, the selected token position may occur before the final answer. The extraction procedure is specified in the repository. Outputs scoring below 0.99 are hypothetically referred for review. We report the proportion flagged and accuracy among unflagged outputs; reviewer decisions and reviewed-system accuracy are not observed.

\subsection{Documentary analysis}

\subsubsection{Selecting the documents}

Documents were collected from an existing UK AI-governance corpus and searches using search engines and official UK and EU websites. We analysed 37 documentary units: 19 issued by UK-level government and public bodies and 18 by EU institutions, bodies and agencies (listed in the appendix). Documents were assessed for current status as of 26 October 2026. The sample was fixed on 31 October 2026, with six unresolved candidates deferred.

The inclusion criteria were that a document (i) was published under the name of a government department, public body, regulator, audit office or the Committee of Public Accounts, or an EU institution, body or agency, including documents written for that body by others; (ii) was relevant to AI used by public officials or in public services; (iii) was treated as current at the cut-off under the selection convention described in the appendix; and (iv) set out recording or disclosure expectations, or how AI outputs should be checked. The study concerns requirements, including non-statutory guidance and recommendations, rather than reports of completed evaluations, audits or deployments. Official web pages were eligible; drafts, consultation versions, blogs and unofficial pages were excluded.

\subsubsection{Coding frame}

Documents were coded using a governance arrangements framework that distinguishes four governance functions: decision-making (who decides whether, when and how AI is used, approved, paused or stopped), delivery (how AI is built, tested, monitored, reviewed and run), coordination (how bodies work with each other and with suppliers) and accountability (who answers for AI, and to whom). Each function is divided into design (the rules, roles and procedures set out) and resources (the staff, skills, guidance, tools, data and money provided or required), producing eight coding cells.

The framework organised passages by governance function and dimension. Measurement theory and management-based regulation informed the interpretation. The unit of coding is a passage that sets out a requirement, provision or expectation bearing on AI used by public bodies. For each passage the coding recorded who holds the requirement, the level of government, its position relative to the state, whether it is statutory, its level of obligation and what the document says follows if it is not met.

\subsubsection{Coding and checking}

Documents were coded using this framework, with AI coding assistance. The author checked every coded entry against the source documents, reviewed the themes and UK/EU interpretations, and corrected identified errors.

\subsubsection{Analysis}

The UK and EU documents were analysed separately and then compared. Within each of the eight function--dimension cells, thematic analysis combined deductive coding, starting from the examples in the coding guidance, with inductive development of additional themes where recurring passages did not fit those examples. This produced 395 subthemes, which were consolidated into 37 themes across the eight cells. Comparison across the eight coding cells produced the documentary findings, which section~4.2 presents by governance function. A table in the appendix reports coding placements, contributing documents and themes.

\subsection{Integrated analysis}

Once both analyses were complete, a literature-informed comparison examined how the sampled requirements address the properties investigated in the experiment and where they leave assessment choices to organisations. It does not replace the documentary coding framework or constitute a prospectively specified assessment instrument. Measurement provides the distinctions between stability, correctness against a reference and robustness of the substantive account; management-based regulation directs attention to how assessment choices and responsibilities are organised. The comparison covers the five checks in Table~\ref{tab:integrated-findings}: repeated runs, configuration comparison, agreement filtering, review routing and robustness of conclusions.

\section{Findings}

This section addresses how UK and EU assurance requirements provide for assessing and justifying the dependability of AI-produced information in public services. Section~4.1 reports what the classification experiment shows about the stability and correctness of individual classifications and the account constructed from them (RQ1, RQ2). Section~4.2 reports what the sampled UK and EU documents ask organisations to establish and how they allocate standard-setting, checking, support, scrutiny and action (RQ3, RQ4). Section~4.3 brings the two studies together.

\subsection{Experimental findings}

The experiment asks whether reasonable changes to how an AI classification task is set up change the results. We compare eight GPT-4o configurations on the same 1,200 Prime Minister's Questions. The anchor is our default configuration, which we also rerun five times without changing its settings. A label agrees with CAP when the model assigns the same topic as the human coders in the Comparative Agendas Project.

Table~\ref{tab:experimental-findings} summarises the six main findings.

\begin{table}[t]
\centering
\caption{Main experimental findings.}
\label{tab:experimental-findings}
\begin{tabular}{@{}p{0.36\linewidth}p{0.57\linewidth}@{}}
\hline
\textbf{Finding} & \textbf{Result} \\
\hline
1. The same setup gives different answers &
Pairs of unchanged runs disagree on 2.78\% of labels on average; 5.42\% of questions change label at least once across five runs. \\[0.5em]
2. Similar accuracy hides different labels &
Accuracy ranges from 68.2\% to 70.3\%, yet 24.3\% of questions receive different labels across configurations. \\[0.5em]
3. Agreement is not correctness &
All eight configurations give the same label that differs from the human coding for 191 questions. \\[0.5em]
4. Dropping disputed questions raises reported accuracy &
Keeping only questions on which all configurations agree raises the anchor's accuracy from 69.6\% to 79.0\%, without changing any labels. \\[0.5em]
5. The same review rule flags very different numbers of outputs &
At the same score threshold, the share flagged for review ranges from 0.9\% to 41.0\%. \\[0.5em]
6. The broad picture holds but the detail shifts &
The same five topics remain most common, but the two largest topics swap order under three configurations. \\
\hline
\end{tabular}
\end{table}

\subsubsection{The same setup gives different answers}

We first checked whether the anchor gives the same answers when its settings remain unchanged. It does not. Across five runs, pairs of runs disagree on 2.78\% of labels on average, and 65 questions (5.42\%) change label at least once. Setting the temperature to zero does not make this pipeline deterministic. We use the 2.78\% figure as a baseline for run-to-run variation.

\subsubsection{Similar accuracy hides different labels}

We then compared the eight configurations. Their accuracy against the human coding is similar, ranging from 68.2\% for free-text output to 70.3\% when the prompt specifies the historical period. The anchor scores 69.6\%. None of the differences in accuracy is statistically significant after correcting for the 28 comparisons.

However, similar accuracy does not mean that the configurations give the same answers. Across the 1,199 questions with labels from all eight configurations, 291 (24.3\%) receive different labels. For 215 questions (17.9\%), the label matches the human coding under some configurations but not others.

\begin{figure}[t]
\centering
\includegraphics[width=0.85\linewidth]{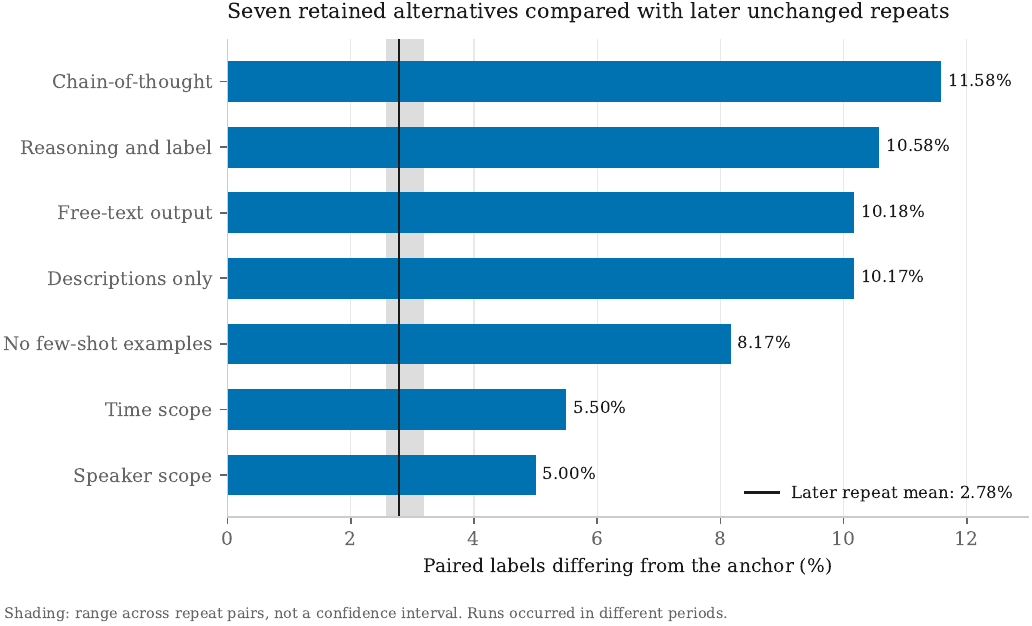}
\caption{Paired label disagreement between each alternative configuration and the anchor. The reference line shows mean disagreement between unchanged repeats; the shaded range shows disagreement across repeat pairs. The comparison spans different execution periods.}
\label{fig:configuration-disagreement}
\end{figure}

Each alternative configuration disagrees with the anchor on between 5.0\% and 11.6\% of labels, compared with the unchanged-repeat baseline of 2.78\% (Figure~\ref{fig:configuration-disagreement}). The repeats were run about five months after the original experiment using an unpinned model alias. Some of this difference may therefore reflect changes over time rather than configuration choices alone.

\subsubsection{Agreement is not correctness}

The human coding provides a reference rather than an unquestionable answer. Some questions cover several topics, as the refuge example in section~4.1.6 illustrates.

The configurations agree with one another more than they agree with the human coders. Because they all use GPT-4o, their agreement is not independent evidence that their labels are right. For 191 questions, all eight give the same label, but that label differs from the human coding. These cases do not establish whether the model or the human coders are right. They show that consistent answers can still disagree with the reference.

\subsubsection{Dropping disputed questions raises reported accuracy}

We next checked what happens to reported accuracy if we keep only questions on which all eight configurations agree. This tests whether selecting cases by agreement changes the apparent performance of the classifier.

The selection leaves 908 of the 1,199 questions. The anchor's accuracy rises from 69.6\% to 79.0\%, an increase of 9.3 percentage points, without a single label changing. Across configurations, the increase ranges from 8.6 to 10.7 points. The higher score describes a selected subset on which performance is better, not an improved classifier.

The exclusions also change the topic mix. For example, half of the Public Lands questions are removed. The accuracy figure therefore needs to be read alongside information about which questions were excluded.

\subsubsection{The same review rule flags very different numbers of outputs}

UK guidance describes sending uncertain outputs to human reviewers (AI\_Insights). We tested a hypothetical version of this: flag any output with an extracted score below 0.99. The procedure extracts the highest probability among valid topic codes at a selected token position.

The same rule produces very different review loads. Under the anchor, 36.0\% of outputs are flagged. Accuracy among those left unflagged is 83.6\%, compared with 69.6\% across all anchor outputs. Under chain-of-thought, only 0.9\% are flagged, and accuracy among unflagged outputs barely differs from the full-sample result: 68.7\% compared with 68.5\%. Across the eight configurations, the share flagged ranges from 0.9\% to 41.0\%.

Under chain-of-thought, the extracted scores are close to one for almost every output, including labels that disagree with the human coding. There is an important limitation: when reasoning precedes the answer, the selected token position may occur before the final classification. The score may therefore not represent confidence in the final label.

These results concern this particular extraction and review procedure. They do not show that reasoning generally makes models more confident, and they do not measure what human reviewers would decide.

\subsubsection{The broad picture holds but the detail shifts}

Topic labels are used to describe what Parliament pays attention to, so we checked whether differences between classifications change that description.

Labels that disagree with the human reference increase the totals for some topics while reducing others. Across the eight configurations, 440 of the 480 labels assigned to the 60 Public Lands questions disagree with the human coding. Of these disagreements, 42.7\% assign International Affairs and 28.2\% assign Government Operations. Compared with the reference, Public Lands therefore looks smaller and those destination topics look larger.

A single question illustrates how different labels emphasise different aspects of a concern. On 8 March 2006, an MP asked the Prime Minister to support women facing violence from their partners, describing a local refuge in financial crisis and women at risk of homelessness. CAP labels it Law and Crime. Five configurations agree, while three label it Social Welfare. Both readings capture part of the question, but a single label records only one. The question stays the same; where it is counted changes.

\begin{figure}[t]
\centering
\includegraphics[width=0.85\linewidth]{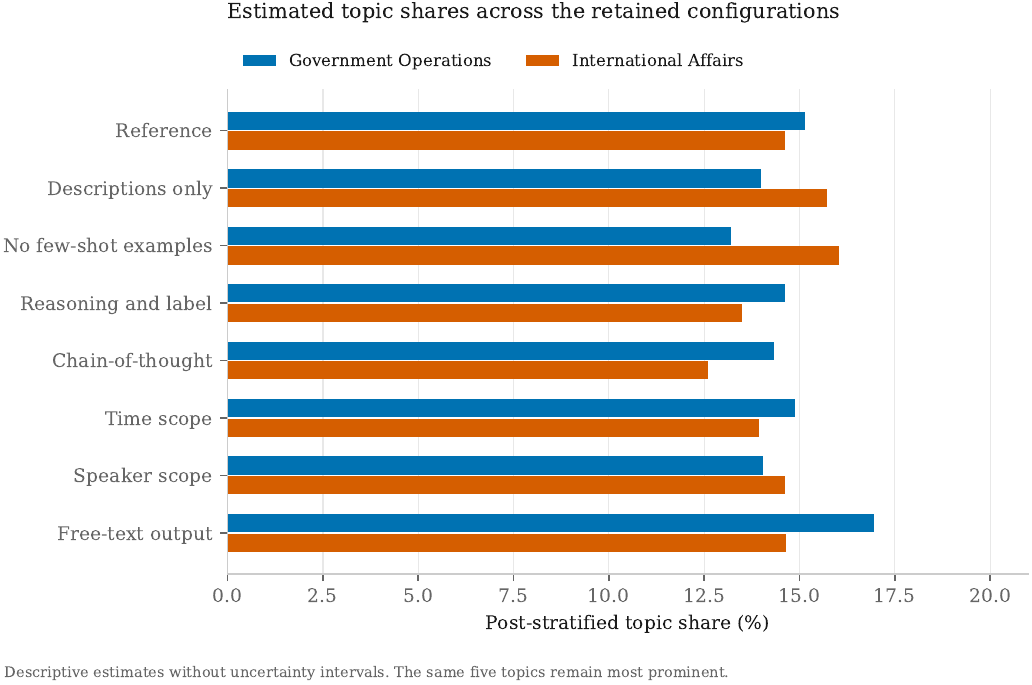}
\caption{Weighted full-period shares of Government Operations and International Affairs across the eight configurations. These descriptive estimates show changes in the size and ordering of the two topics.}
\label{fig:topic-shares}
\end{figure}

We then compared topic shares across the full period, weighting them to account for the stratified sample. The same five topics are most common under every configuration, but their shares and order vary (Figure~\ref{fig:topic-shares}). Government Operations accounts for between 13.2\% and 17.0\% of questions, and International Affairs for between 12.6\% and 16.0\%. Under three configurations, International Affairs exceeds Government Operations. This reversal does not occur in any of the five unchanged repeats.

\subsection{Documentary findings}

We examined what the UK and EU documents expect organisations to check about AI outputs and who is responsible for assessment, support, scrutiny and action. The findings concern the requirements set out in the documents, rather than how they work in practice. Document codes refer to the appendix list. Under the AI Act, a provider develops an AI system, or has one developed, and places it on the market or puts it into service under its own name; a deployer uses it; high-risk systems include those listed in Annex III. The findings are organised by the four governance functions of the coding framework (decision-making, delivery, coordination and accountability), each examined through its design and resources (section~3.2). Table~\ref{tab:documentary-findings} in the appendix summarises the findings by function.

\subsubsection{Delivery: organisations retain important choices about testing}

In both the UK and EU documents, AI outputs are required, expected or recommended to be tested before use, monitored, re-tested on change and reviewed by people where the stakes are high. The documents leave important choices about references, metrics and acceptance thresholds to providers and deploying bodies, subject to applicable duties. The AI Act requires purpose-appropriate, predefined metrics and probabilistic thresholds for testing high-risk systems (EU046, Art.~9(8)). The UK government's AI Playbook asks bodies to identify objective measures for assessing outputs, without prescribing a common metric or threshold (UK174). The AI testing framework gives an illustrative threshold of more than 80\% of sampled outputs judged acceptable by reviewers (UK071).

Stability is addressed at different levels of detail. The AI Act requires testing to ensure consistent performance for the intended purpose, without specifying a repeated-run agreement statistic in the cited provision (EU046, Art.~9(6)). UK guidance names particular checks: the audit offices' guide recommends running the same prompt several times (UK268), and the testing framework recommends checking for inconsistent or unexplained outputs from similar prompts (UK071). Neither passage prescribes a common agreement statistic or a specific response to each disagreement.

Oversight expectations go beyond appointing a reviewer. In the UK, the Playbook calls for validation of high-risk decisions and meaningful intervention (UK174), and AI Insights describes routing uncertain or sensitive outputs to trained reviewers. The testing framework recommends testing handover triggers and override mechanisms, and whether non-technical users can identify hallucinations, bias or off-topic outputs (UK071). In the EU, Article~14 addresses effective oversight, anomaly detection, automation bias, interpretation, override and intervention, while Article~26(2) requires competence, training, authority and support. Organisations retain important choices about how review is implemented and evaluated.

\subsubsection{Decision-making, coordination and resources: responsibility sits in different places}

The UK documents emphasise the responsibilities of bodies using AI, while the EU documents combine substantial provider duties with deployer responsibilities. The Playbook tells bodies to adopt a risk-based approach and consider whether AI is appropriate for high-risk applications (UK174). Under the AI Act, deployers of relevant high-risk systems must follow instructions for use, and specified public-body uses also carry registration and fundamental-rights impact-assessment duties (EU046, Arts~26, 27, 49).

A public body that builds a classifier on a bought-in model may itself be the provider: the definition includes a public authority putting a system into service under its own name (Art.~3(3)). Provider and deployer responsibilities therefore do not necessarily sit with separate organisations. The contrast between the UK and EU documents also partly reflects their different levels and scope, rather than an exhaustive difference between the two legal systems.

What a body learns about a tool also differs. The UK transparency template allows the model specification to be left blank for some off-the-shelf models (UK087), while the AI Act requires instructions for use to declare accuracy levels and metrics (EU046, Arts~13, 15(3)).

External assessment and approval depend on the system and its use. For the relevant Annex III systems, Article~43(2) provides for internal-control conformity assessment, in which providers assess their own systems. Other categories can involve notified bodies or sectoral assessment routes. Assessing whether a system meets requirements is distinct from independently authorising a public body's particular use. For example, independent authorisation applies to specified police biometric uses. UK guidance also describes internal approval, including possible senior or ethics-board approval for high-risk AI (UK071). These arrangements do not establish a universal independent approval process for ordinary classification.

Resource duties coexist with shared support. UK documents identify training, guidance and support for regulator capability (UK174; UK139). The AI Act provides for training and advice for deployers and requires adequate resources for national competent authorities (Arts~62, 70(3)). The documents identify support arrangements but do not establish whether organisations can access sufficient staff, time or expertise to carry out particular checks.

\subsubsection{Accountability: consequences and accountability vary}

The UK and EU documents set out consequences and routes for scrutiny, but these differ in how binding they are and who they apply to. UK examples include disciplinary action for DWP non-compliance (UK190) and potential legal action under equality law (UK216). These are consequences for breaches of particular duties, rather than automatic penalties for an incorrect AI output.

The EU documents include statutory enforcement and fines, with Member States determining the extent of administrative fines on their public authorities and a separate regime for Union institutions (EU046, Arts~99(8), 100). The AI Act also requires deployers to suspend use under specified risk conditions (Art.~26(5)).

Both sets of documents contain public-facing and supervisory routes. UK transparency records are to name an accountable role rather than an individual (UK086) and describe review or, where relevant, challenge arrangements (UK087). The EU combines notifications and documentation for supervisors with public database information, subject to specified exceptions, complaints (Art.~85) and a scoped right to explanation (Art.~86).

\subsection{Integrated findings}

Table~\ref{tab:integrated-findings} compares the experimental findings with the relevant requirements and identifies the assessment choices that remain.

\begin{table*}[t]
\centering
\caption{Experimental checks, relevant requirements and remaining assessment choices.}
\label{tab:integrated-findings}
\begin{tabular}{@{}p{0.21\textwidth}p{0.27\textwidth}p{0.43\textwidth}@{}}
\hline
\textbf{Check} & \textbf{Experimental finding} & \textbf{Relevant requirements and remaining choices} \\
\hline
Repeated runs (Table~\ref{tab:experimental-findings}, finding~1) &
Unchanged runs disagree on 2.78\% of labels on average. &
UK guidance recommends identical- and similar-prompt checks; the AI Act requires consistency. The cited provisions do not prescribe a common repeated-run agreement statistic. \\[0.5em]
Configuration comparison (Table~\ref{tab:experimental-findings}, findings~2 and~3) &
Similar accuracy accompanies different labels and shared disagreements with the human reference. &
Guidance addresses model comparison and evaluation. The cited passages do not consistently specify reporting of changes in individual labels or shared disagreements with the reference. \\[0.5em]
Agreement filtering (Table~\ref{tab:experimental-findings}, finding~4) &
Excluding disputed questions raises reported accuracy without changing any label. &
The cited passages do not specifically address this selection effect. \\[0.5em]
Score-based review routing (Table~\ref{tab:experimental-findings}, finding~5) &
Between 0.9\% and 41.0\% of outputs are flagged at the same threshold. &
Requirements address meaningful oversight, uncertainty and reviewer capability. How the task-specific score is extracted and used still needs examination. \\[0.5em]
Robustness of conclusions (Table~\ref{tab:experimental-findings}, finding~6) &
The same five topics remain most common, while their shares and order change. &
Risk, impact and intended-purpose assessment are relevant. The specific analysis of topic shares and rankings is not prescribed. \\
\hline
\end{tabular}
\end{table*}

The comparison shows that relevant checks are already recognised. UK guidance explicitly addresses repeated prompts and model comparisons, while the AI Act requires purpose-appropriate testing and consistent performance for high-risk systems. However, these expectations do not consistently specify how organisations should report the particular differences found in the experiment.

Human review illustrates why the implementation details matter. Requirements for competent reviewers and meaningful intervention depend partly on which outputs reach those reviewers. In our experiment, the same threshold flagged very different proportions of outputs across configurations, subject to the score-extraction limitation. Stating that a human reviews uncertain outputs therefore leaves a further question: does the procedure identify the outputs that need attention?

Together, the studies show why choices about evaluation need justification. Configuration, filtering and score extraction changed classifications, reported accuracy and hypothetical review coverage. The documents place assessment choices within duties, guidance and scrutiny arrangements, but do not consistently specify reporting of those effects. This does not establish that broader duties exclude these checks or that an assessment omitting them would comply with all applicable requirements.

\section{Discussion}

The experiment shows that configurations with almost the same accuracy gave different labels to about a quarter of questions, and that evaluation choices changed reported accuracy, the share of outputs sent for review and the account of political attention. The documents show that UK and EU requirements recognise relevant checks but leave important choices about measures, thresholds and reporting to providers and public bodies. This matters particularly in public services, where public bodies need to justify their use of AI-produced evidence, and that justification should be open to challenge by those affected.

\subsection{Why AI-produced evidence needs scrutiny}

Information can shape which issues receive attention, how a situation is understood and which responses appear justified before anyone makes a final decision. Several of the provisions identified in section~4.2 attach to decisions or to specific uses: the right to explanation applies to specified decisions (Art.~86), and independent approval applies to specified uses such as police biometric identification. These use-specific provisions do not by themselves explain how the dependability of evidence used to inform decisions should be assessed. The European Commission identifies text mining to support responses to parliamentary questions as a potential use~\cite[section~3.2]{ec2024aiatec}. Our experiment shows that classification choices can change the account of parliamentary attention.

This informational role makes representation an assurance concern. Classification leaves the source statement unchanged but changes where it appears in an organised account. A statement raising several concerns may be recorded under only one topic, and alternative classifications can foreground different aspects. In Parliament, the relationship between wording and response is explicit: a minister's answer ``should be confined to the points contained in the question''~\cite[para.~19.16]{erskinemay1916}.

The significance of these findings depends on what people rely on AI-produced information for. Different classifications may offer useful alternative readings during exploration, although this study does not test that benefit. Variation that is tolerable during exploration may require closer examination when outputs contribute to service assessments, resource allocation, policy priorities or scrutiny. The issue is not variation itself but whether it is visible and accounted for when outputs are used as evidence.

\subsection{What this adds to existing work}

Research on LLM evaluation has shown that results depend on prompts and repeated runs~\cite{mizrahi2024state,barrie2026prompt,yorke2026behind}. Work on predictive multiplicity has shown that models with similar performance can disagree on individual cases~\cite{marx2020predictive,damour2022underspecification}. Our findings are consistent with this work and extend it by following variation through to reported evaluation results, the conclusions drawn from outputs and the procedure for routing outputs to review. \citet{green2022flaws} questions the assumptions behind human oversight; our routing result shows that which outputs reach a reviewer can itself depend on configuration and score extraction. \citet{frohnapfel2026multiplicity} connect multiplicity to the AI Act's accuracy provisions. We add an experiment and a comparison of UK and EU requirements, showing where these recognise relevant checks and where they leave assessment choices to organisations. Measurement shows why stability and correctness must be checked separately. Management-based regulation helps interpret this allocation of discretion and directs attention to how choices are justified, supported and scrutinised.

\subsection{What evidence should support discretion}

Allowing organisations to choose measures and thresholds can help them assess AI for their particular tasks. Different public-service uses may need different standards. Those choices nevertheless require evidence about properties that an aggregate performance score cannot establish.

For consequential informational uses, guidance could ask organisations to report variation across unchanged repeats and plausible configurations, identify shared errors and evaluation exclusions, and examine whether the conclusions drawn from outputs remain robust. Where confidence scores trigger human review, the extraction and routing procedure should also be examined. These are candidate checks motivated by this study, to be selected and extended according to the task. They supplement existing guidance rather than constitute a complete assurance framework: safety, fairness, privacy, security, accessibility and reviewer effectiveness require their own appropriate assessment.

Public bodies commissioning or using AI should ask for evidence of how configuration choices, evaluation exclusions and review procedures affect the results they intend to rely on. Providers should supply the information needed to examine these effects.

A short assurance record could identify the intended use and why error or variation matters; the tested model version and configuration; the evaluation population and exclusions; aggregate results alongside relevant individual-label changes, shared disagreements with the reference and changes in substantive conclusions; and the review rule, responsible assessor and triggers for further action. Untested properties should be stated explicitly. This is a proposal for supporting scrutiny, rather than a reporting intervention whose effectiveness the study has tested.

\subsection{Human review and responsibility}

Human oversight depends partly on which outputs reach reviewers, not simply on appointing competent reviewers. Our routing result illustrates this implementation issue, although reviewer effectiveness was not tested. Providers and deploying bodies should record who justifies the assessment, who can scrutinise it and what triggers action; a public body may itself hold both roles.

\section{Limitations}

The experiment is exploratory: it covers one task, one model and a small set of configurations, and more extensive testing is needed to estimate how large these differences are in general. Correctness means agreement with human-coded CAP labels, which are themselves judgements and may be wrong in some cases. The configurations come from one model family, so their agreement with one another is not independent evidence. Labels and extracted scores were retained for eight of the original 17 configurations. The unchanged repeats were run about five months later on an unpinned model alias, so comparisons do not isolate configuration effects from change over time. Extracted scores describe the implemented routing procedure rather than validated final-answer confidence, and reviewer decisions were not observed. The experiment does not establish effects on actual public-service decisions or harm.

Documents were coded using the study's governance framework, with AI assistance in applying it, and the author checked every coded entry; no inter-coder agreement rate is reported. The document sample concerns formal requirements, excludes Member States' national guidance and does not establish actual compliance, capacity, enforcement or effectiveness. The UK and EU documents also operate at different levels of government and with different scopes.

Further work should compare configurations and unchanged repeats within the same execution period using pinned model versions, validate final-answer score extraction and test actual reviewer decisions. Studies within public bodies should examine whether output differences change the evidence reported to decision-makers and how assurance requirements are applied in practice. Summarisation, assessments and advice require investigation on their own terms.

\section{Conclusion}

The dependability of AI-produced information in public services is both an empirical and an institutional question. It is empirical because stability, correctness against a reference and robustness of conclusions must be examined rather than inferred from aggregate performance. It is institutional because organisations must decide which properties matter, what variation is acceptable, who performs the assessment and how its justification can be challenged. Public bodies need guidance on selecting relevant checks, the capability to perform them and clear responsibility for justifying reliance on the results. The contribution of this study is to connect what technical checks establish with who must justify reliance on AI-produced evidence and how that justification can be scrutinised.

\section*{Generative AI usage statement}

The documents were coded using the study's coding framework, with assistance from AI coding assistants. AI coding assistants also helped with analysis code, checking and editing. The author checked all AI-assisted work and is responsible for the paper.

\section*{Reproducibility and data availability}

Data and code are available at
\url{https://github.com/Yorkel/cap-assurance-release}.

Configuration specifications, prompts, statistical details, the score-extraction procedure and detailed results are available in the repository.

\bibliographystyle{ACM-Reference-Format}
\bibliography{references}

\clearpage
\appendix
\raggedbottom
\section{Configuration dimensions}

\begin{table}[H]
\centering
\caption{Experimental configuration dimensions.}
\label{tab:configuration-dimensions}
\begin{tabular}{@{}p{0.17\linewidth}p{0.76\linewidth}@{}}
\hline
\textbf{Dimension} & \textbf{Anchor and alternatives} \\
\hline
Codebook & Full descriptions, rule and six examples; descriptions only; descriptions and rule without examples \\
Reasoning & Label only; reasoning then label; explicit step-by-step reasoning then label \\
Scope & No scope; historical period; speaker's party \\
Output & Requested JSON; free text; provider JSON mode; strict function schema \\
Decoding & One temperature-zero sample; temperature 0.7; five-sample self-consistency; 20-token limit; label logit bias \\
Examples & Original examples; different topics; different exemplars from the original topics \\
Provider & GPT-4o; Claude Sonnet under the anchor instructions \\
\hline
\end{tabular}
\end{table}

\section{Documentary findings by governance function}

\begin{table}[H]
\centering
\caption{Main documentary findings by governance function.}
\label{tab:documentary-findings}
\begin{tabular}{@{}p{0.13\textwidth}p{0.17\textwidth}p{0.30\textwidth}p{0.31\textwidth}@{}}
\hline
\textbf{Function} & \textbf{Question} & \textbf{UK documents} & \textbf{EU documents} \\
\hline
Decision-making &
Whether to use AI &
Each body applies the test to itself &
The deployer decides; some practices are banned (Art.~5) \\[0.5em]
&
Approval before use &
Internal approval, plus central spend approvals and ICO consultation where data-protection risk remains &
Independent authorisation for police biometric identification; third-party conformity assessment for some systems (Art.~43(1)) \\[0.5em]
&
Pausing or stopping a tool &
Named in guidance rather than binding rules &
The deployer must suspend use if it has reason to consider a risk (Art.~26(5)); authorities can order withdrawal or recall (Art.~79(2)) \\[0.5em]
Delivery &
What to measure, and the pass mark &
Bodies choose measures within applicable duties; illustrative thresholds in guidance &
The provider (Art.~9(8)) \\[0.5em]
&
Stability over repeated runs and setups &
Non-binding guidance only &
Consistency required (Arts~9(6), 15(1)); no repeated-run statistic named \\[0.5em]
&
Which outputs a person reviews &
Depends on use and uncertainty; guidance includes handover and reviewer-capability tests &
Risk and context shape oversight (Art.~14) \\[0.5em]
Coordination &
What the body learns about the tool &
Transparency record; model version may be left blank for off-the-shelf tools &
Instructions for use declare accuracy levels and metrics (Arts~13, 15(3)) \\[0.5em]
Accountability &
Who is responsible for outputs &
The body that uses the tool, even if bought in &
The deployer for its use; the provider for the system's conformity \\[0.5em]
&
What follows a failure &
Seldom specified for a failed check; disciplinary action, legal action under equality law &
Member States set fines for public bodies; separate regime for Union institutions \\[0.5em]
Resources (all functions) &
Staff, money and time &
Training, guidance and regulator support identified; actual capacity not measured &
``Adequate'' resources for authorities, unquantified \\
\hline
\end{tabular}
\end{table}

\clearpage
\section{Documentary sample and coding}

\subsection{The 37 documents}

\begin{longtable}{@{}p{0.19\textwidth}p{0.43\textwidth}p{0.25\textwidth}p{0.06\textwidth}@{}}
\caption{The 37 documents in the sample (19 UK, 18 EU), with the year of the version analysed.}
\label{tab:documentary-sample}\\
\hline
\textbf{ID} & \textbf{Title} & \textbf{Issued by} & \textbf{Year} \\
\hline
\endfirsthead

\multicolumn{4}{l}{\tablename~\thetable\ continued}\\
\hline
\textbf{ID} & \textbf{Title} & \textbf{Issued by} & \textbf{Year} \\
\hline
\endhead

\hline
\multicolumn{4}{r}{Continued on next page}\\
\endfoot

\hline
\endlastfoot

UK010 &
Ethical AI Use in the Energy Sector (version 2) &
Ofgem &
2026 \\[0.4em]

UK071 &
AI Testing and Assurance Framework for Public Sector &
Cross-Government Testing Community &
2025 \\[0.4em]

UK082 &
AI Opportunities Action Plan &
Matt Clifford for DSIT &
2025 \\[0.4em]

UK085 &
ATRS: Mandatory Scope and Exemptions Policy &
Government Digital Service &
2024 \\[0.4em]

UK086 &
ATRS: Guidance for Public Sector Bodies &
Government Digital Service &
2025 \\[0.4em]

UK087 &
ATRS Template, version 4.0 &
Government Digital Service &
n.d. \\[0.4em]

UK099 &
Data and AI Ethics Framework &
Government Digital Service &
2025 \\[0.4em]

UK133 &
Guidance on the Impact Evaluation of AI Interventions &
HM Treasury, Evaluation Task Force &
2025 \\[0.4em]

UK136 &
Guidelines for AI Procurement &
Office for Artificial Intelligence &
2020 \\[0.4em]

UK139 &
Implementing the UK's AI Regulatory Principles: Initial Guidance for Regulators &
DSIT &
2024 \\[0.4em]

UK174 &
Artificial Intelligence Playbook for the UK Government &
Government Digital Service &
2025 \\[0.4em]

UK190 &
Artificial Intelligence Security Policy &
Department for Work and Pensions &
2026 \\[0.4em]

UK206 &
Introduction to AI Assurance &
DSIT &
2024 \\[0.4em]

UK207 &
The Roadmap to an Effective AI Assurance Ecosystem &
Centre for Data Ethics and Innovation &
2021 \\[0.4em]

UK216 &
AI in Public Services; AI: Checklist for Public Bodies in England &
Equality and Human Rights Commission &
2022 \\[0.4em]

UK266 &
Good Practice Guide for Organisations Using AI &
National Audit Office &
2026 \\[0.4em]

UK268 &
Auditing AI Systems: A Guide for Public Auditors &
Six supreme audit institutions, including the UK National Audit Office &
2025 \\[0.4em]

NAO\_\allowbreak AI\_\allowbreak Strategy &
The NAO's AI Strategy &
National Audit Office &
2026 \\[0.4em]

AI\_\allowbreak Insights &
AI Insights (15 chapters) &
Government Digital Service &
2026 \\[0.4em]

EU046 &
Artificial Intelligence Act, Regulation (EU) 2024/1689 &
European Parliament and Council &
2024 \\[0.4em]

EU\_\allowbreak Omnibus &
Regulation (EU) 2026/1744 amending the AI Act (Digital Omnibus on AI) &
European Parliament and Council &
2026 \\[0.4em]

EU043 &
Guidelines on Prohibited AI Practices &
European Commission &
2025 \\[0.4em]

EU088 &
Guidelines on the Obligations of Providers of General-Purpose AI Models &
European Commission &
2025 \\[0.4em]

EU090 &
Explanatory Notice and Template for the Public Summary of Training Content &
European Commission (AI Office) &
2025 \\[0.4em]

EU\_\allowbreak Art50 &
Guidelines on the Article 50 Transparency Obligations &
European Commission &
2026 \\[0.4em]

EU121 &
Code of Practice on Transparency of AI-Generated Content &
European Commission (publisher) &
2026 \\[0.4em]

AIatEC &
Artificial Intelligence in the European Commission (AI@EC) &
European Commission &
2024 \\[0.4em]

EC\_\allowbreak staff\_\allowbreak GenAI\_\allowbreak 2023 &
Guidelines for Staff on Online Generative AI Tools &
European Commission &
2023 \\[0.4em]

EP2020 &
Guidelines for the Responsible Use of AI in the European Parliament &
European Parliament &
2020 \\[0.4em]

EP2024 &
Artificial Intelligence: A Governance Framework &
European Parliament &
2024 \\[0.4em]

EDPS\_\allowbreak AI\_\allowbreak risk &
Guidance for Risk Management of AI Systems &
European Data Protection Supervisor &
2025 \\[0.4em]

EDPS\_\allowbreak GenAI\_\allowbreak orientations &
Generative AI and the EUDPR: Orientations (version 2) &
European Data Protection Supervisor &
2025 \\[0.4em]

EU074 &
Opinion 28/2024 on Data Protection Aspects of AI Models &
European Data Protection Board &
2024 \\[0.4em]

ENISA\_\allowbreak frontier &
ENISA's View on Cybersecurity in the Frontier AI Era &
EU Agency for Cybersecurity &
2026 \\[0.4em]

JRC129100 &
AI Watch: Road to the Adoption of AI by the Public Sector &
European Commission, Joint Research Centre &
2022 \\[0.4em]

JRC143539 &
Advancing AI Adoption in EU Public Administrations &
European Commission, Joint Research Centre &
2026 \\[0.4em]

ECA\_\allowbreak strategy &
Artificial Intelligence Strategy 2026--2030 &
European Court of Auditors &
2026 \\

\end{longtable}

Versions analysed date from 2020 to 2026; 30 of the 37 date from 2024--2026 (UK 15 of 19, EU 15 of 18). One UK document (UK087) is undated.

\subsection{Document selection}

The selection convention treated an official source without a withdrawal, archive or supersession notice at the cut-off as current; this is not a finding that every legal provision was already applicable. One author-approved exception was the accessible 2023 Commission staff guidance, substituted for the unavailable 3 July 2025 version and identified as the older text rather than verified current guidance. The EU090 text analysed is the Explanatory Notice and Template, C(2025) 8311 final, dated 5 December 2025; its preserved source filename carries an earlier July date. LLM-assisted screening recorded supporting passages; the hand-edited selection table and sample freeze govern membership. AI Insights' 15 chapters count as one unit. Related supporting components are retained within their mapped unit; EU088 and EU090 are separate units despite sharing one selection-table row. The 37 units correspond to 54 mapped source-text files.

\subsection{Coding counts}

Coding counts are reported in Table~\ref{tab:coding-counts}.

Documents denotes documentary units with at least one placement in the cell.

\begin{table}[H]
\centering
\caption{Documentary coding placements, contributing documents and themes.}
\label{tab:coding-counts}
\begin{tabular}{@{}p{0.29\linewidth}rrrrr@{}}
\hline
\textbf{Cell} &
\shortstack{\textbf{UK coding}\\\textbf{placements}} &
\shortstack{\textbf{UK}\\\textbf{documents}} &
\shortstack{\textbf{EU coding}\\\textbf{placements}} &
\shortstack{\textbf{EU}\\\textbf{documents}} &
\textbf{Themes} \\
\hline
Decision-making: design & 362 & 19 & 614 & 18 & 6 \\
Decision-making: resources & 52 & 18 & 45 & 15 & 4 \\
Delivery: design & 666 & 19 & 521 & 18 & 7 \\
Delivery: resources & 257 & 19 & 259 & 18 & 4 \\
Coordination: design & 227 & 19 & 289 & 18 & 5 \\
Coordination: resources & 65 & 19 & 60 & 14 & 3 \\
Accountability: design & 354 & 19 & 463 & 17 & 5 \\
Accountability: resources & 74 & 17 & 67 & 15 & 3 \\
\hline
Total & 2,057 & 19 & 2,318 & 18 & 37 \\
\hline
\end{tabular}
\end{table}

Placements count a passage in every cell to which it was assigned, so the 4,375 placements (UK 2,057; EU 2,318) are not unique passages. The 37 themes were built from 395 subthemes developed separately for the UK and EU documents.

\end{document}